# Observation of spin-polarized bands and domain-dependent Fermi arcs in polar Weyl semimetal MoTe$_2$

M. Sakano[1,2], M. S. Bahramy[1,3], H. Tsuji[1], I. Araya[1], K. Ikeura[1], H. Sakai[1,4], S. Ishiwata[1,5], K. Yaji[2], K. Kuroda[2], A. Harasawa[2], S. Shin[2], and K. Ishizaka[1,3]

[1]*Quantum-Phase Electronics Center (QPEC) and Department of Applied Physics, The University of Tokyo, Tokyo 113-8656, Japan*
[2]*Institute for Solid State Physics, The University of Tokyo, Kashiwa, Chiba 277-8581, Japan*
[3] *RIKEN Center for Emergent Matter Science (CEMS), Wako 351-0198, Japan*
[4]*Department of Physics, Osaka University, Toyonaka, Osaka 560-0043, Japan*
[5]*PRESTO, Japan Science and Technology Agency (JST), Tokyo 102-8666, Japan*

## Abstract

We investigate the surface electronic structures of polar 1$T$"-MoTe$_2$, the Weyl semimetal candidate realized through the nonpolar-polar structural phase transition, by utilizing the laser angle-resolved photoemission spectroscopy combined with first-principles calculations. Two kinds of domains with different surface band dispersions are observed from a single-crystalline sample. The spin-resolved measurements further reveal that the spin polarizations of the surface and the bulk-derived states show the different domain-dependences, indicating the opposite bulk polarity. For both domains, some segment-like band features resembling the Fermi arcs are clearly observed. The patterns of the arcs present the marked contrast between the two domains, respectively agreeing well with the slab calculation of (0 0 1) and (0 0 -1) surfaces. The present result strongly suggests that the Fermi arc connects the identical pair of Weyl nodes on one side of the polar crystal surface, whereas it connects between the different pairs of Weyl nodes on the other side.

There has been increasing interest in Weyl semimetals as a new class of the topological state of matter, where the nondegenerate bands form pairs of band crossings (i.e. Weyl nodes) at the Fermi level ($E_F$). The bulk electronic structure of Weyl semimetals is characterized by the spin-polarized Weyl cone dispersions formed through the breaking of either time-reversal or space-inversion symmetry [1-4]. At the surface, on the other hand, the chiral charge associated with the Weyl nodes warrants the existence of the gapless surface states, so-called Fermi arcs that connect the two-dimensionally (2D) projected Weyl nodes. Due to these unusual bulk and surface electronic states, a variety of new magnetoelectric phenomena have been predicted [4-9]. Until now, several experimental verifications of realistic Weyl semimetal compounds have been raised (eg. the TaAs family [10-12]). Recently, theoretical studies also predicted that $1T'$-$M$Te$_2$ ($M$ = Mo, W) is a candidate type-II Weyl semimetal with canted Weyl cones, where the nodes appear as the intersections of the electron and hole Fermi surfaces [13-16]. The realization of type-II Weyl semimetal is highly desired from the viewpoint of exploring a new topological phase and its peculiar low-energy excitations [13].

$1T'$-MoTe$_2$ has a CdI$_2$-type structure consisting of an edge-sharing MoTe$_6$ octahedral network, which is strongly distorted by forming Mo-Mo bonding chains. It shows a structural transition from monoclinic ($P2_1/m$) to orthorhombic ($Pnm2_1$) on cooling through 250 K [17], which accompanies the nonpolar-polar transition. Consequently, the low-temperature phase, so-called $T_d$-MoTe$_2$ is classified as an unusual polar semimetal. This phase transition is being discussed in relation to the appearance of the superconductivity [18] as well as the critical enhancement of Seebeck effect [19, 20]. A number of recent angle-resolved photoemission spectroscopy (ARPES) studies on MoTe$_2$ reported the semimetal structure and signature of Fermi arcs, suggestive of type-II Weyl semimetal [21-27]. While several studies deeply discuss the significance of Fermi arcs with regard to the type-II Weyl semimetal state by comparing with a number of numerical results [22, 27], the experimental demonstration of the polarity induced spin-polarized bulk band, as well as the clear comparison of Fermi arcs for top and bottom surfaces, in the light of their connectivity, are yet unsatisfactory.

In this work, we investigate $1T'$-MoTe$_2$ by means of high-resolution laser ARPES. The spin polarization of the surface and bulk-derived bands for respective domains are confirmed by the spin-resolved laser ARPES (laser-SARPES). By raising the temperature to 100 K, we successfully find the signature of Fermi arcs that reside about 30 meV above $E_F$. The results are in good agreement with the band calculation, and indicate that the Fermi arcs connect the identical pair of W1 nodes at the (0 0 -1) surface, whereas they connect the different pairs of the W1 and W2 nodes at the (0 0 1) surface.

The single-crystalline 1T'-MoTe$_2$ was synthesized as reported elsewhere [19, 20]. ARPES at 25 K was performed using the He-discharge lamp (21.2 eV) and the fourth harmonic generation of Ti:sapphire laser (6.43 eV, *s*-polarized light) [28], with a VG-Scienta R4000WAL analyzer. The total energy resolution was set to 10 and 3 meV, respectively. For the laser-SARPES at 25 K and the ARPES at 100 K, a 6.99 eV laser (*s*-polarized light) and a ScientaOmicron DA30L analyzer mounted with two sets of very-low-energy electron diffraction spin detectors were used at the Institute of Solid State Physics (ISSP), The University of Tokyo [29]. The total energy resolution was set to 30 and 3 meV, respectively. Samples were cleaved *in situ* at room temperature. All measurements were performed in ultrahigh vacuum better than $1 \times 10^{-10}$ Torr. Electronic structure calculations were performed within the context of density functional theory (DFT) using the Perdew–Burke–Ernzerhof exchange-correlation functional as implemented in the VASP program [30, 31]. Relativistic effects were fully included. The structure parameters in Ref. [14] were used, and the corresponding Brillouin zone was sampled by a $20 \times 10 \times 5$ *k*-mesh. For the orbital and layer projection calculation, a tight-binding supercell containing 80 layers of MoTe$_2$ was constructed by downfolding the DFT results using maximally localized Wannier functions [32], employing Mo 4*d* orbitals and Te 5*p* orbitals as the projection centers.

Figure 1 shows the band structures obtained by ARPES and the calculations. The Brillouin zone for the low temperature orthorhombic structure is depicted in Fig. 1(a) with the momentum axes ($k_x$, $k_y$, $k_z$). The ARPES image obtained by the HeIα (21.2 eV) photon is shown in Fig. 1(c). Along Γ-X, the complicated hybridization of bands forming the semimetallic hole and electron pockets are recognized near $E_F$, whereas rather simple holelike band dispersions are observed along Γ-Y. The band calculations along Γ-X focusing on the near-$E_F$ semimetallic bands are shown in Figs. 1(f) and 1(g), respectively, for the monoclinic (non-polar) high-temperature and the orthorhombic (polar) low-temperature phases. In the low-temperature phase, due to the polarity along the stacking direction (*c*-axis), the spin-splitting of the bands occurs *via* the spin-orbit coupling (inside the green broken curves) [33]. These spin-polarized bands can be regarded as the hallmark of the polar semimetal state that produces the Weyl nodes located at $E - E_F = 6$ meV (W1) and 59 meV (W2) at ($k_x$, $k_y$, $k_z$) = (0.185, 0.013, 0) and (0.181, 0.053, 0) Å$^{-1}$ as reported in Ref. [14].

To focus on the near-$E_F$ electronic structure, the ARPES images along Γ-X recorded by using a 6.43 eV laser are shown in Figs. 1(d) and (e). Owing to the small spot size (120 μm × 100 μm), we always find two types of domains [surfaces A and B, respectively shown in Figs. 1(d) and 1(e)], by surveying the positions on the sample surface [Fig. 1(b)]. In the present study, we assign these as the polar domains with the opposite bulk polarity (*vide infra*). By comparing with the slab calculations in Figs. 1(h) and 1(i), we can recognize the rather flat surface band (S1, green curve). It is located at around $E - E_F$ = -80 meV (-50 meV) for Surface A (B), which merges with the bulk hole band at $k_x$ ~ 0.2 Å$^{-1}$. The other surface band (S2, red curve) has a dispersive character which crosses with the S1 bands. These S1 and S2 bands seem to connect the bulk electron and hole pockets, indicating their topologically protected nature. The band features reflecting the bulk electron pocket (brown curves), on the other hand, are similarly observed in surfaces A and B. From this viewpoint, the slab calculations for the (0 0 1) and (0 0 -1) surfaces, as shown in Figs. 1(h) and 1(i), show qualitatively good agreements with the bands observed for Surface A and B, respectively [34]. By taking advantage of these domains, the termination dependence of the topological Weyl state can be investigated.

Here we use laser-SARPES (6.99 eV) to finely resolve the spin-polarized band structures along $k_x$. The ARPES intensities of spin up and spin down for y-spin ($s_y$) components are plotted in Figs. 2(c) and 2(d), respectively. They show marked differences, indicating that there are multiple bands with the opposite $s_y$ polarizations. The energy distribution curves for $s_y$-up (red) and $s_y$-down (blue) are displayed in Fig. 2(e), with the peak positions depicted by circular markers. By plotting these peaks, as shown in Figs. 2(c) and 2(d), we can trace the band dispersions with respective spin polarizations. The guides for these spin-polarized bands are overlaid on the high *k*-resolution laser (6.43 eV) ARPES image in Fig. 2(a). Here the red curves indicate the bands with $s_y$-up components, whereas the blue / light-blue curves represent those with $s_y$-down. From this result, we can safely say that the S1 (S2) surface state has the $s_y$-up ($s_y$-down) polarization. We also note that there is some characteristic near-$E_F$ intensity with $s_y$-up polarization observed at around $k_x$ = 0.25 – 0.3 Å$^{-1}$ (detector angle $\theta$ = 1° – 5°), depicted by the light-blue curves and ○ markers in Figs. 2(a), and 2(d) and (e).

The results for surface B are similarly shown in Figs. 2(f)-2(j). By comparing with surface A, we can derive several features regarding the spin-polarized bands. The surface states S1 and S2 apparently have the same spin polarization for both surfaces, whereas the near-$E_F$ intensity has the opposite spin polarization ($s_y$-down for surface A and $s_y$-up for surface B). Here we note that the spin polarization for the bulk bands which split by the spin-orbit coupling, as presented in Fig. 1(g), should be opposite when we compare the polar domains. Taking these into account, we assign the bands depicted by light-blue and pink ○ markers and curves in Figs. 2(e) and 2(j) and Figs. 2(a) and 2(f), respectively, as those derived from the bulk band components. These arguments, with the help of the slab calculations [Figs. 2(b) and (g)], strongly suggest that surfaces A and B should correspond to the (0 0 1) and (0 0 -1) surfaces, respectively.

To seek for the possible Fermi arcs connecting the Weyl nodes located off the Γ-X line, we measured the energy contours for surfaces A (0 0 1) and B (0 0 -1), as respectively shown in Figs. 3(a)-3(d) and 3(e)-3(h). The high $k$-resolution data at $E_F$, recorded at 25 K with a 6.43 eV laser, present some clear differences for respective surfaces [Figs. 3(a) and (e)]. For surface A (0 0 1), we observe a pair of segment-like features [indicated by blue markers in Fig. 3(e)] touching the circular contour surrounding the bulk electron pocket, whereas no such clear segments are recognized for surface B (0 0 -1) [Fig. 3(a)]. The pair of segments observed in surface A (0 0 1) may be the tails of the Fermi arcs connecting W1 and W2 nodes, as predicted in the previous calculations [14]. To get closer to the W1 node in energy ($E = E_F + 6.7$ meV), we performed the measurement at 100 K with a 6.99 eV laser. Owing to the thermal excitation, we can detect the electrons up to 30 meV above $E_F$, as shown in Figs. 3(b)-3(d) and 3(f)-3(h). At $E - E_F = 30$ meV [Fig. 3(h)], we find a segment-like feature in surface B (0 0 -1) that is centered at $k_y = 0$ (indicated by the blue marker), which makes a clear contrast to the one observed in surface A (0 0 1). The segment observed for surface B (0 0 -1) seems to well correspond to the Fermi arc connecting the pair of W1 nodes across the Γ-X line, in contrast to those connecting the W1 and W2 nodes in Surface A (0 0 1). This difference for respective surfaces is also well reproduced in the slab calculations as shown in Figs. 3(i)-3(k) and 3(m)-3(o), respectively. The schematic of the energy contours at W1 ($E_F + 6.7$ meV) is indicated in Figs. 3(l) and 3(p), where the brown shaded areas represent the 2D projected bulk bands and the blue curves indicate the surface states [35].

Since the W1 and W2 nodes have very close $k_x$ values ($k_x \sim 0.19$ Å$^{-1}$), the constant-$k_x$ (i.e., $E$ vs $k_y$ plane) is a good cut to show the dispersion relations of the multiple Weyl cones and related Fermi arcs. The $E$-$k_y$ ARPES images for surfaces A (0 0 1) and B (0 0 -1) recorded at 100 K are shown in Figs. 4(b) and 4(e), respectively. The positions of $k_x$ are indicated by the blue broken lines in Figs. 4(a) and 4(d). By looking at the ARPES images divided by the Fermi-Dirac distribution function of 100 K, we see a clear contrast between the two surfaces. For surface B (0 0 -1), there is a high intensity region at $E - E_F > 30$ meV, $k_y \sim 0$, indicated by the blue broken curve. This should correspond to the Fermi arc centered at $k_y = 0$ connecting two W1 nodes, as shown in the calculation in Fig. 4(c). For surface A (0 0 1), on the other hand, there are symmetric dispersions appearing above the Fermi level at around $k_y \sim 0.05$ Å$^{-1}$ (blue broken curves). They should be the signatures of the disconnected Fermi arcs connecting the W1 and W2 nodes. These dispersions of the Fermi arcs are also well presented in the slab calculation in Fig. 4(f).

The present results thus strongly suggest the surface-orientation-dependent Fermi arcs indeed realized in polar MoTe$_2$. In non-centrosymmetric Weyl semimetals, different patterns of Fermi arcs on the top and bottom surfaces are theoretically expected [36, 37]. Indeed, an ARPES study on NbP [38] revealed the nonequivalent forms of Fermi arcs on respective surfaces, that are terminated at the same 2D projected bulk Weyl nodes. Also for MoTe$_2$, the difference of the surface states and Fermi arcs for both terminations are discussed in comparison with calculations [22, 27]. The present result on MoTe$_2$ indicates the different connectivity of the Fermi arcs to the two kinds of Weyl nodes for (0 0 1) and (0 0 -1) terminated surfaces, as predicted in Ref. [36]. However, an additional Fermi arc that connects the pair of W2 nodes on surface B (0 0 -1) should exist for consistently connecting the Weyl nodes. Since the W2 nodes are expected to reside at 59 meV above $E_F$, which is hardly accessible at < 200 K, more direct investigation of the unoccupied state should be required to fully understand the whole picture of the Fermi arcs [21].

In conclusion, the laser (spin-resolved) ARPES study on 1$T$'-MoTe$_2$ revealed the two kinds of domains with different surface band dispersions. The spin polarizations of these surfaces states and the bulk-derived electronic structures showed the different domain dependences, reflecting the opposite bulk polarity of the crystal structure. We further found that the signature of Fermi arcs appearing in the ARPES energy contours possess the nonequivalent forms for both domains, respectively connecting the identical pair of W1 nodes (domain B) and the different pairs of W1 and W2 nodes (domain A). They agree well with the slab calculations on the (0 0 -1) and (0 0 1) surfaces, respectively. The present results demonstrate a credible support for the type-II Weyl semimetal state realized in the polar 1$T$'-MoTe$_2$.

Acknowledgement

This research was partly supported by the Photon Frontier Network Program of the MEXT; Research Hub for Advanced Nano Characterization, The University of Tokyo, supported by MEXT, Japan; and Grant-in-Aid for Scientific Research from JSPS, Japan (KAKENHI 15H03683 and 16K133815).

Reference

1. S. Murakami, New J. Phys. **9**, 356 (2007).
2. X. Wan, A. M. Turner, A. Vishwanath, and S. Y. Savrasov, Phys. Rev. B **83**, 205101 (2011).
3. A. A. Burkov and L. Balents, Phys. Rev. Lett. **107**, 127205 (2011).
4. G. Xu, H. Weng, Z. Wang, X. Dai, and Z. Fang, Phys. Rev. Lett. **107**, 186806 (2011).
5. H. B. Nielsen and M. Ninomiya, Phys. Lett. B **130**, 389 (1983).
6. A. A. Zyuzin and A. A. Burkov, Phys. Rev. B **86**, 115133 (2012).
7. Z. Wang and S.-C. Zhang, Phys. Rev. B **87**, 161107(R) (2013).
8. M. M. Vazifeh and M. Franz, Phys. Rev. Lett. **111**, 027201 (2013).
9. A. A. Burkov, J. Phys.: Condens. Matter **27**, 113201 (2015).
10. S.-Y. Xu, I. Belopolski, N. Alidoust, M. Neupane, G. Bian, C. Zhang, R. Sankar, G. Chang, Z. Yuan, C.-C. Lee, S.-M. Huang, H. Zheng, J. Ma, D. S. Sanchez, B. Wang, A. Bansil, F. Chou, P. P. Shibayev, H. Lin, S. Jia, and M. Z. Hasan, Science **349**, 613 (2015).
11. B. Q. Lv, H. M. Weng, B. B. Fu, X. P. Wang, H. Miao, J. Ma, P. Richard, X. C. Huang, L. X. Zhao, G. F. Chen, Z. Fang, X. Dai, T. Qian, and H. Ding, Phys. Rev. X **5**, 031013 (2015).
12. L. X. Yang, Z. K. Liu, Y. Sun, H. Peng, H. F. Yang, T. Zhang, B. Zhou, Y. Zhang, Y. F. Guo, M. Rahn, D. Prabhakaran, Z. Hussain, S.-K. Mo, C. Felser, B. Yan, and Y. L. Chen, Nat. Phys. **11**, 728-732 (2015).
13. A. A. Soluyanov, D. Gresch, Z. Wang, Q. S. Wu, M. Troyer, X. Dai, and B. A. Bernevig, Nature (London) **527**, 495 (2015).
14. Y. Sun, S.-C. Wu, M. N. Ali, C. Felser, and B. Yan, Phys. Rev. B **92**, 161107(R) (2015).
15. T.-R. Chang, S.-Y. Xu, G. Chang, C.-C. Lee, S.-M. Huang, B. Wang, G. Bian, H. Zheng, D. S. Sanchez, I. Belopolski, N. Alidoust, M. Neupane, A. Bansil. H.-T. Jeng, H. Lin and M. Z. Hasan, Nat. Commun. **7**, 10639 (2016).
16. Z. Wang, D. Gresch, A. A. Soluyanov, W. Xie, S. Kushwaha, X. Dai, M. Troyer, R. J. Cava, and B. A. Bernevig, Phys. Rev. Lett, **117**, 056805 (2016).
17. R. Clarke, E. Marseglia, and H. P. Hughes, Philos. Mag. **38**, 121 (1978).
18. Y. Qi, P. G. Naumov, M. N. Ali, C. R. Rajamathi, Y. Sun, C. Shekhar, S.-C. Wu, V. S

üß, M. Schmidt, E. Pippel, P. Werner, R. Hillebrand, T. Förster, E. Kampertt, W. Schnelle, S. Parkin, R. J. Cava, C. Felser, B. Yan, and S. A. Medvedev, Nat. Commun. **7**, 11038 (2016).

19. K. Ikeura, H. Sakai, M. S. Bahramy, and S. Ishiwata, APL Materials. **3**, 041514 (2015).

20. H. Sakai, K. Ikeura, M. S. Bahramy, N. Ogawa, D. Hashizume, J. Fujioka, Y. Tokura, and S. Ishiwata, Sci. Adv. **2**, e1501117 (2016).

21. I. Belopolski, S.-Y. Xu, Y. Ishida, X. Pan, P. Yu, D. S. Sanchez, M. Neupane, N. Alidoust, G. Chang, T.-R. Chang, Y. Wu, G. Bian, S.-M. Huang, C.-C. Lee, D. Mou, L. Huang, Y. Song, B. Wang, G. Wang, Y.-W. Yeh, N. Yao, J. E. Rault, P. Le Fevre, F. Bertran, H.-T. Jeng, T. Kondo, A. Kaminski, H. Lin, Z. Liu, F. Song, S. Shin, and M. Z. Hasan, Phys. Rev. B **94**, 085127 (2016).

22. L. Huang, T. M. McCormick, M. Ochi, M.-T. Suzuki, R. Arita, Y. Wu, D. Mou, H. Cao, J. Yan, N. Trivedi, and A. Kaminski, Nat. Mater. **15**, 1155-1160 (2016).

23. K. Deng, G. Wan, P. Deng, K. Zhang, S. Ding, E. Wang, M. Yan, H. Huang, H. Zhang, Z. Xu, J. Denlinger, A. Fedorov, H. Yang, W. Duan, H. Yao, Y. Wu, Y. S. Fan, H. Zhang, X. Chen, and S. Zhou, Nat. Phys. **12**, 1105 (2016).

24. J. Jiang, Z. K. Liu, Y. Sun, H. F. Yang, R. Rajamathi, Y. P. Qi, C. Yang, L X adn Chen, H. Peng, C.-C. Hwang, S. Z. Sun, M. S.-K., I. Vobornik, J. Fujii, S. S. P. Parkin, C. Felser, B. H. Yan, and Y. L. Chen, Nat. Commun. **8**, 13973 (2017).

25. A. Liang, J. Huang, S. Nie, Y. Ding, Q. Gao, C. Hu, S. He, Y. Zhang, C. Wang, B. Shen, J. Liu, P. Ai, L. Yu, X. Sun, W. Zhao, S. Lv, D. Liu, C. Li, Y. Zhang, Y. Hu, Y. Xu, L. Zhao, G. Liu, Z. Mao, X. Jia, F. Zhang, S. Zhang, F. Yang, Z. Wang, Q. Peng, H. Weng, X. Dai, Z. Fang, Z. Xu, C. Chen, and X. J. Zhou, arXiv:1604.01706.

26. N. Xu, Z. J. Wang, A. P. Weber, A. Magrez, P. Bugnon, H. Berger, B. B. Fu, B. Q. Lv, N. C. Plumb, M. Radovic, K. Conder, T. Qian, J. H. Dil, J. Mesot, H. Ding, and M. Shi, arXiv:1604.02116.

27. A. Tamai, Q. S. Wu, I. Cucchi, F. Y. Bruno, S. Riccò, T.K. Kim, M. Hoesch, C. Barreteau, E. Giannini, C. Besnard, A. A. Soluyanov, and F. Baumberger, Fermi Arcs and Their Topological Character in the Candidate Type-II Weyl Semimetal $MoTe_2$, Phys. Rev. X **6**, 031021(2016).

28. Y. Suzuki, T. Shimojima, T. Sonobe, A. Nakamura, M. Sakano, H. Tsuji, J. Omachi, K.

Yoshioka, M. Kuwata-Gonokami, T. Watashige, R. Kobayashi, S. Kasahara, T. Shibauchi, Y. Matsuda, Y. Yamakawa, H. Kontani, and K. Ishizaka, Phys. Rev. B **92**, 205117 (2015).

29. K. Yaji, A. Harasawa, K. Kuroda, S. Toyohisa, M. Nakayama, Y. Ishida, A. Fukushima, S. Watanabe, C. Chen, F. Komori, and S. Shin, Rev. Sci. Instrum. **87**, 053111 (2016).

30. J. P. Perdew, K. Burke, and M. Ernzerhof, Phys. Rev. Lett. **77**, 3868 (1996).

31. G. Kresse *et al.*, http://cms.mpi.univie.ac.at/vasp/ .

32. A. A. Mostofi, J. R. Yates, Y. -S. Lee, D. Vanderbilt, and N. Marzari, Comput. Phys. Commun. **178**, 685 (2008).

33. K. Ishizaka, M. S. Bahramy, H. Murakawa, M. Sakano, T. Shimojima, T. Sonobe, K. Koizumi, S. Shin, H. Miyahara, A. Kimura, K. Miyamoto, T. Okuda, H. Namatame, M. Taniguchi, R. Arita, N. Nagaosa, K. Kobayashi, Y. Murakami, R. Kumai, Y. Kaneko, Y. Onose, and Y. Tokura, Nat. Mater. **10**, 521 (2011).

34. By comparing the energy of the bottom of the bulk electron pocket, we can estimate that the Fermi level for the ARPES result is about 30 meV lower as compared to the band calculation. It should be attributed to the nonstoichiometry of the single crystalline samples where the excess holes tend to be introduced by the Te difficiency [20].

35. One may notice that the W1-W1 Fermi arc in surface A (0 0 1) appears better at higher energy ($E > E_F + 15$ meV), whereas the W1-W2 Fermi arcs in surface B (0 0 -1) are more distinguishable at lower energy ($E < E_F + 15$ meV). It can be explained by the difference of the distributions of the Fermi arcs in the energy vs 2D momentum space. The W1-W1 Fermi arc in surface A (0 0 1) quickly disappears at the energy below the W1 nodes, by merging to the surface state of circular contour. The W1-W2 Fermi arcs, on the other hand, can be seen surviving to the lower energies below $E_F$, thus making it easier to observe by low temperature ARPES.

36. R. Okugawa and S. Murakami, Phys. Rev. B **89**, 235315 (2014).

37. S.-M. Huang, S.-Y. Xu, I. Belopolski, C.-C. Lee, G. Chang, B. Wang, N. Alidoust, G. Bian, M. Neupane, A. Bansil, H. Lin, and M. Z. Hasan, Nat. Commun. **6**, 7373 (2015).

38. S. Souma, Zhiwei Wang, H. Kotaka, T. Sato, K. Nakayama, Y. Tanaka, H. Kimizuka, T. Takahashi, K. Yamauchi, T. Oguchi, K. Segawa, and Y. Ando, Phys. Rev. B **93** 161112(R) (2016).

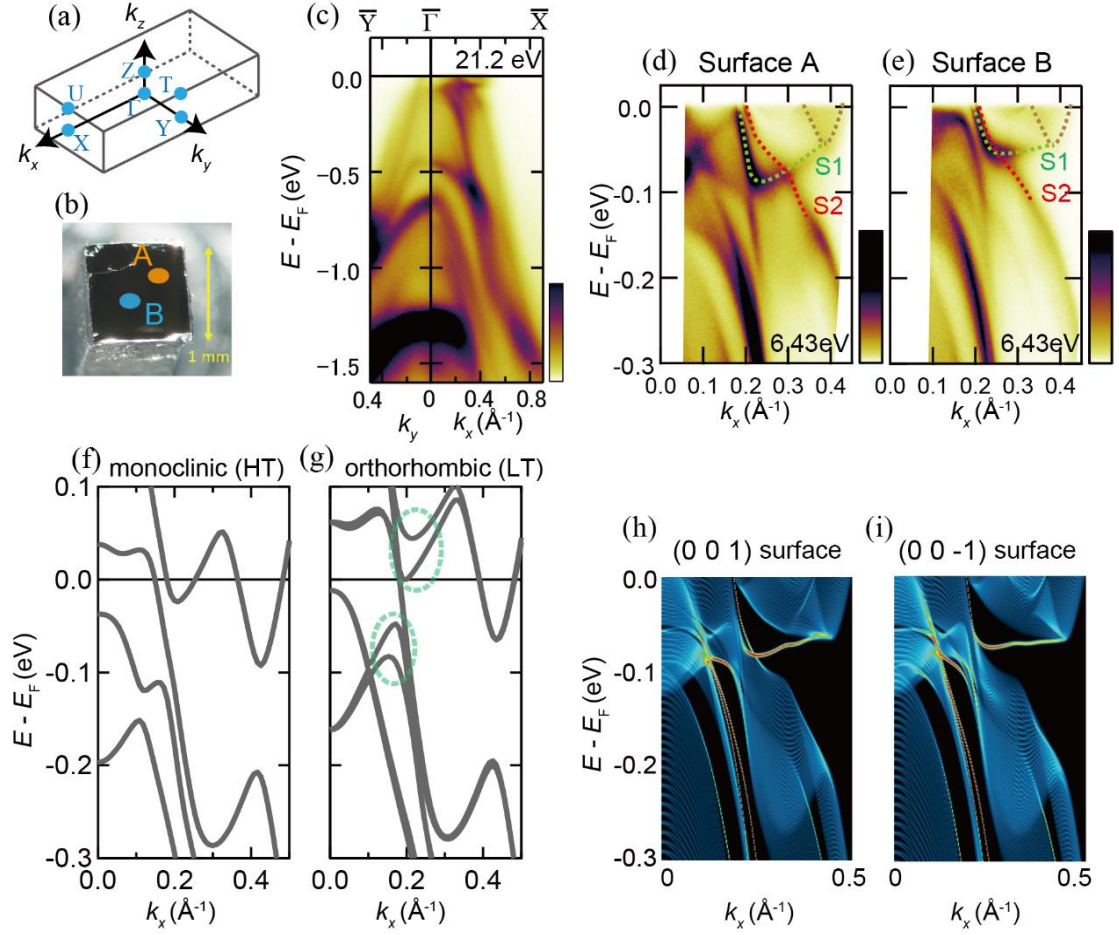

**FIG. 1.** (a) The Brillouin zone of orthorhombic $MoTe_2$. (b) The typical sample of single-crystalline $MoTe_2$. The markers A and B show the spots where the two different kinds of the surface states (namely, surface A and B) are observed. (c) ARPES image recorded at 25 K with the photon energy of 21.2 eV. (d),(e) ARPES images along the $k_x$ axis recorded at 25 K with a 6.43 eV laser, from surfaces A and B, respectively. The green (red) broken curves indicate the S1 (S2) surface states, whereas the brown broken curves denote the bulk-derived bands. (f),(g) The band calculations of $MoTe_2$ along ($k_x$, 0, 0) for high-temperature (HT) monoclinic and low-temperature (LT) orthorhombic structures, respectively. The green broken curves in (g) indicate the regions where the spin splitting is induced by the polar structure. (h),(i) The slab calculation along the $k_x$ axis for the (0 0 1) and (0 0 -1) surfaces, respectively.
10

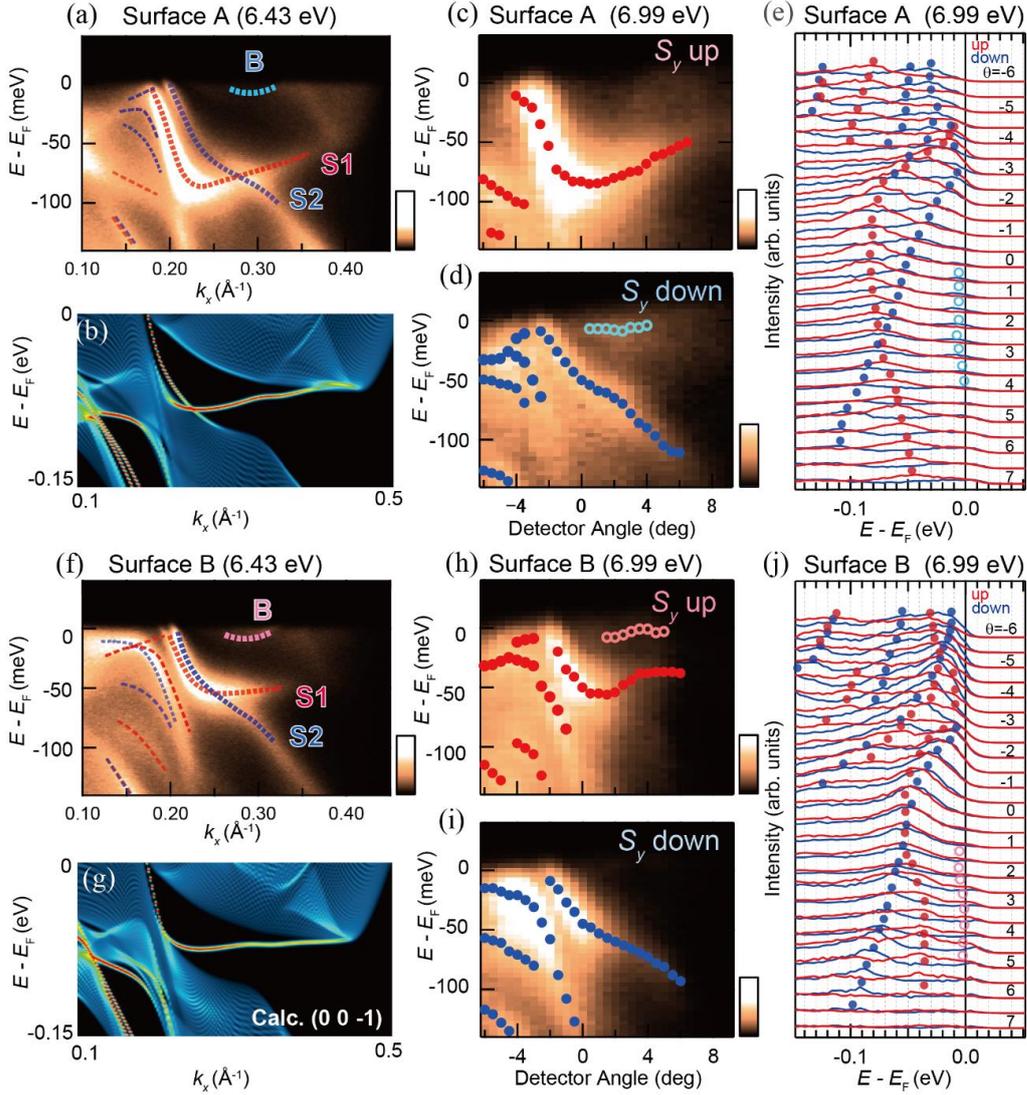

**FIG. 2,** (a) ARPES image along the $k_x$ axis for surface A obtained by a 6.43 eV laser. The blue (red/pink) broken curves indicate the guides for the band dispersions with $s_y$-down ($s_y$-up) polarizations. (b) The corresponding slab calculation for the (0 0 1) surface. (c),(d) Laser-SARPES images for $s_y$-up and $s_y$-down, respectively, obtained from surface A by using a 6.99 eV laser (ISSP). The detector slit is fixed along the $k_x$ axis. (e) Energy distribution curves (EDCs) for $s_y$-up (red) and $s_y$-down (blue), respectively. The markers indicate the positions of the intensity peaks, which are also plotted in (c),(d). (f)-(j) A similar set of ARPES image, slab calculation, laser-SARPES image, and EDC data is presented for surface B and the (0 0 -1) surface.

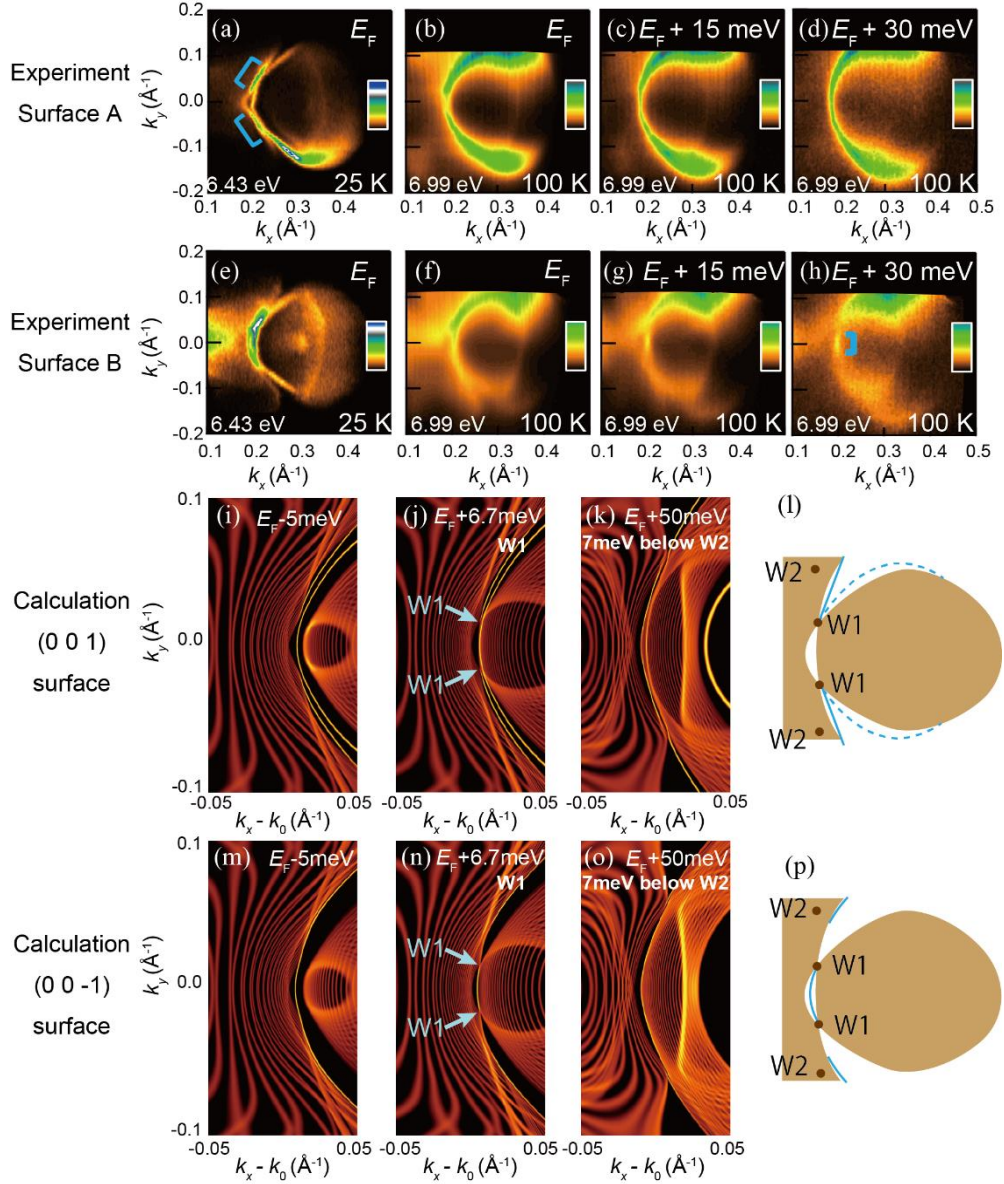

**FIG. 3.** The energy contours at $E_F$ recorded at 25 K (a) and $E_F$, $E_F + 15$ meV, $E_F + 30$ meV recorded at 100 K [(b)-(d)], respectively, obtained from surface A. The similar set of energy contours from surface B is shown in (e)-(h). The data at 25 K (100 K) are obtained using a 6.43 eV laser (6.99 eV laser at ISSP). The calculated energy contours at $E_F - 5$ meV, $E_F + 6.7$ meV (i.e., W1), and $E_F + 50$ meV (i.e., 7 meV below W2) are shown in (i)-(k), respectively, for the (0 0 1) surface. The arrow markers in (j) indicate W1 nodes. The corresponding schematic energy contour at the W1 energy is depicted in (l), with the brown shades indicating the 2D projected bulk and the blue curves indicating the surface states. The similar set of data for the (0 0 -1) surface is shown in (m)-(p).

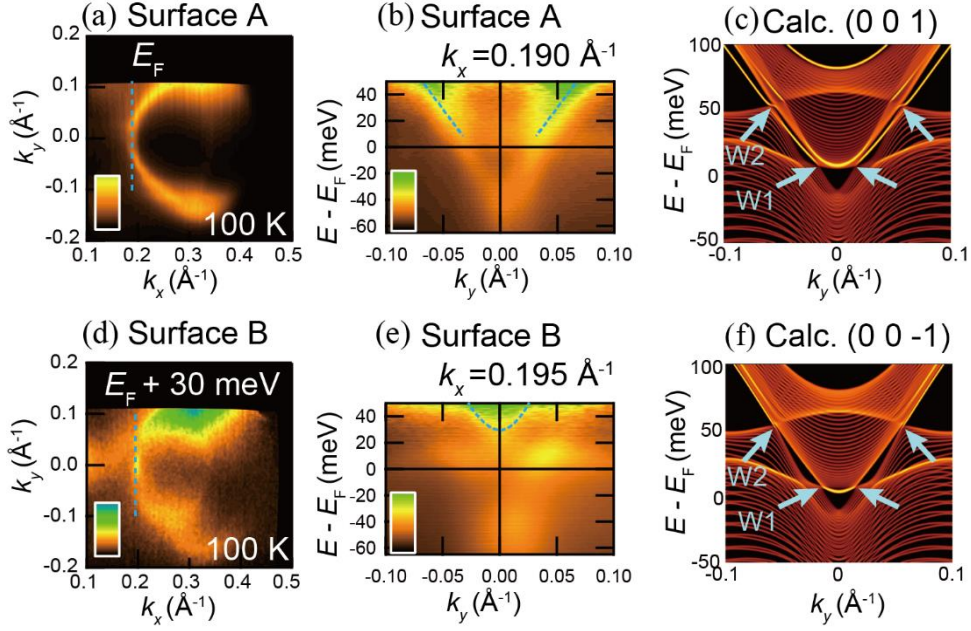

**FIG. 4.** The energy contour at $E_F+30$ meV (a) and $E$-$k_y$ ARPES image divided by the Fermi Dirac distribution function at 100 K (b), and the calculated $E$-$k_y$ image at $k_x = k_0$ (c), obtained for the (0 0 1) surface. The energy contour at $E_F$ (d) and $E$-$k_y$ ARPES image divided by the Fermi-Dirac distribution function at 100 K (e), and the calculated $E$-$k_y$ image at $k_x = k_0$ (f), obtained for the (0 0 -1) surface. The arrow markers in (c) and (f) indicate the W1 and W2 nodes. The ARPES data are obtained using a 6.99 eV laser at ISSP.